\documentclass[aps,prb,twocolumn,amsmath,amssymb,groupedaddress,superscriptaddress,longbibliography]{revtex4-1}
\usepackage{hyperref}
\usepackage{epsfig}
\usepackage{color}
\usepackage{graphicx}
\usepackage{bm}
\usepackage{epstopdf}
\usepackage{amsmath}
\usepackage{tikz}
\usepackage{todonotes}
\usepackage{subfigure}
\usepackage{textcomp}
\usepackage{wasysym}
\usepackage{general_latex}
\usepackage{hyperref}
\newcommand{\angstrom}{\textup{\AA}}

\begin{document}
\title{Exact surface-wave spectrum of a dilute quantum liquid}

\author{Peter V. Pikhitsa}
\affiliation{Department of Mechanical and Aerospace Engineering,
Seoul National University, 08826 Seoul, Korea}
\author{Uwe R. Fischer}
\affiliation{Center for Theoretical Physics, Department of Physics and Astronomy, Seoul National University, 08826 Seoul, Korea}

\begin{abstract}
We consider a dilute gas of bosons with repulsive contact interactions,
described on the mean-field level by the Gross-Pitaevski\v\i\/ equation, and {bounded} by an impenetrable ``hard'' wall (either rigid or flexible). We solve the Bogoliubov-de Gennes equations for excitations on top of the Bose-Einstein condensate analytically, by using matrix-valued hypergeometric functions. This leads to the exact spectrum of gapless Bogoliubov excitations localized near the boundary.
The dispersion relation for the surface excitations represents for small wavenumbers $k$ a ripplon mode with  fractional power law dispersion for  a flexible wall, and a phonon mode (linear dispersion) for a rigid wall. For both types of excitation  we provide, for the first time, the exact dispersion relations of the dilute quantum liquid for all $k$ along the surface, extending to
$k \rightarrow \infty$.
The small wavelength excitations are shown to be bound to the surface with a maximal
binding energy $\Delta= \frac18 (\sqrt{17}-3)^2 mc^2 \simeq 0.158\, mc^2$, which both 
types of excitation  asymptotically approach, where $m$ is mass of bosons and $c$ bulk speed of sound.
We demonstrate that  this binding energy is close to the experimental value obtained 
for surface excitations of helium II confined in nanopores, 
reported in \href{https://doi.org/10.1103/PhysRevB.88.014521}{Phys. Rev. B {\bf 88}, 014521 (2013).}

\end{abstract}


\date{\today}

\maketitle

\section{Introduction}
Initially, the Gross-Pitaevski\v\i\/ equation (GPE) was intended
as a model to describe structures and excitations in superfluid helium.\cite{Gross1961,Pitaevskii} Being a nonlinear Schr\"{o}dinger equation, it was however recognized later on that it possesses a variety of applications for various nonlinear processes in condensed matter such as bright and dark solitons in dilute Bose-Einstein condensates (BECs, for which the GPE is accurate on the mean-field level)\cite{Pitaevskii1} and nonlinear optics,\cite{Carusotto}, as well as finite amplitude waves on the surface of a liquid.\cite{Zakharov} Excitations on top of the mean-field ground state representing the BEC,  known as Bogoliubov excitations,\cite{Bogoliubov} 
are described by the eigenmodes of the matrix Bogoliubov-de Gennes  equations (BdGE).
The associated quanta of the perturbation field
have become the archetype of quasiparticle excitations
in superconductivity \cite{Bogoliubov1958,Valatin1958,deGennes} and the theory of dilute quantum gases,\cite{LeggettRMP}, {\it inter alia} also for the formulation of the propagation of quantum fields
on effective curved spacetimes.\cite{Kurita}
The ubiquitous nature of the BdGE makes rigorous analytical solutions highly desirable,
but very few, and only in limiting cases, have been obtained.

Domain wall solutions of the GPE such as 2D dark solitons are known to be unstable except for those in the presence of a hard wall. However the case of a hard wall deserves investigation in particular because it is connected with the generic topic of edge excitations in topological phases.
Specifically, the corresponding physical situation bears some resemblance to two-band models with Majorana bound states that arise as solutions to a BdG approach. The gapless modes that propagate along a physical boundary, while they are exponentially decaying
away from the physical boundary, are gapless boundary modes or edge states.\cite{Hasan}

Examples for the occurrence of surface excitations in {bounded} BECs comprise, for example, superfluid $^4\!$He (helium II) confined in pores,\cite{Shams} self-bound condensates at the low-density surface of superfluid helium \cite{Griffin}, as well as surface states of a BEC trapped in an external potential   \cite{Anglin}, or surface states of other media with a defocusing nonlinearity.\cite{Kuznetsov} 
They are of fundamental interest since they reveal the role of quantum effects on the excitation character (i.e., effects which are not existing on the classical level) in restricted geometries.  

Considering the   boundary condition of a hard wall  
for the surface of a trapped BEC, the stability of surface bound states was examined in \cite{Kuznetsov}, by imposing  that the wave function vanishes at the wall. 
The corresponding surface potentials, much steeper than harmonic, have been prepared by using laser sheets to trap the dilute quantum gas (for example, in \cite{Gaunt}).
An inhomogeneous stationary solution of the GPE (the ``domain wall") which coincides with the half of the dark soliton (kink) at rest,\cite{Pitaevskii1}
may have as one of its physical realizations a hard wall \cite{Kuznetsov} where localized Bogoliubov excitations were proposed to exist.\cite{Pikhitsa}  However, the full analytical solution for the corresponding surface-bound 
excitations has not been found before.At large wavelengths, one class of these excitations represents a surface phonon and the other a ripplon.
Our approach is inherently quantum, as it operates near the node plane of the
domain wall-soliton, and is hence based on an inherently nonclassical (vector-valued)
wavefunction, and is not restricted to large wavelengths, where the (essentially quantum) kinetic terms are small. We note that the existence of
a short-wavelength surface excitation (a ``surface roton") was previously conjectured,\cite{Reut}  but its possible connection to capillary waves was then stated as being doubtful.
We will see below that for both classes of excitations, starting either from surface phonon or ripplon at large wavelengths, small-wavelength surface excitations exist, with a binding energy approached by both types of excitation  at large momenta.

The hard wall boundary condition approximates the steepness of  the effective potential at the  free surface of liquid helium, which was proven to be composed of a nearly pure condensate of dilute bosonic gas that satisfies the GPE.\cite{Griffin}
{The wave function of the BEC is a quantum order parameter that approximately describes the condensate in real liquid helium below the superfluid transition. The helium background (including a well-defined surface) fixes the natural boundary conditions for the BEC. Therefore, the BEC concept accomodates both liquid helium II and a dilute superfluid Bose gas bounded by an external wall.}  

One may consider the free kink wall with profile $\psi_{0}=\tanh(x)$ extending into the bulk of the liquid $(x\ge 0)$ to model the free surface, 
demanding only the topological stability of such a solution for which its nodal surface undergoes weak flexural oscillations. Then the position of the hard wall is flexible (like an impenetrable membrane on the surface of helium II) and imitates the free surface of the liquid. The  liquid surface of helium II is under these provisos equivalent to a hard wall container.

Here we consider the problem of localized gapless excitation modes by finding analytical solutions of a matrix Schr\"{o}dinger equation which we show to be equivalent to the BdGE.\cite{Chen,Kuznetsov,Muryshev,Pikhitsa}
While recently, Ref.~\cite{Nambu} obtained such an analytical solution in the presence of a domain wall,  it is restricted to large wavelengths, and furthermore faces the difficulty of extrapolation to the case of an infinite-size surface.
{We stress that even the classical ripplon (fractional power law) spectrum at small wavenumbers is not  trivially obtained from the BdGE, where no classical (phenomenological) surface tension is assumed \emph{a priori}. In a BEC, the surface tension itself is expressed using Planck's constant and thus is of an inherently quantum nature.}

The binding energy of localized excitations is a primary quantity of interest. Recent experiments that prove the common physical origin of the Landau description of a superfluid and the BEC description \cite{Diallo} support the view that the binding energy is relevant. Furthermore,  neutron scattering experiments in helium II \cite{Prisk} reveal a surface excitation that directly gives the binding energy.
Remarkably, we show that the spectrum of surface excitations can be calculated analytically for {\em any} wavevector $k$, reproducing the numerical results and with the analytical results obtained for the  limiting cases $k \rightarrow 0$ and $k \rightarrow \infty$.
We have solved the BdGE for the case of the domain wall (see Eqs.~(4.16-4.19) in \cite{Nambu}). The 
limit of $k \rightarrow \infty$, which in the bulk BEC results in the energy spectrum $\varepsilon = {\hbar^2 k^2}/ {2m}+\mu$ where $m$ is the mass of the boson and $\mu=gn_0$ is the chemical potential while $g$ and $n_0$ are the coupling constant and the BEC particle density, respectively, then leads to $\varepsilon = {\hbar^2 k^2}/ {2m}+\mu - \Delta$.

\section{Bogoliubov-de Gennes equations}
\subsection{Basic setup}
The GPE of a scalar quantum gas
can be written as:\cite{Ginzburg1}
\bea
i\hbar\frac{\partial \psi}{\partial t}= - \frac{\hbar^2}{2m}\nabla^2 \psi + gn_0(|\psi|^2-1)\psi.
\label{eq:s0}
\ea
We introduce dimensionless quantities by measuring distances in units of the healing length $\xi=\hbar/mc$ and energies in units of  the ``rest mass energy" $gn_0=mc^2$ where $c=\sqrt{gn_0/m}$ is the sound velocity.
The stationary version of Eq.~(\ref{eq:s0}) for a kink with node at the position $x=0$ gives
the wavefunction $\psi_0=\tanh(x)$ of the soliton.
We will impose perturbations on this solution to investigate its Bogoliubov excitations by representing $\psi$ of Eq.~(\ref{eq:s0}) as a sum of plane waves:\cite{Pitaevskii}
 $
 \psi = \psi_0(x) + \vartheta( \vec{r}, t )$ with $\vartheta( \vec{r}, t ) = a_{\omega,\vec k}(x) \exp (i \vec{k}\cdot \vec{\varrho} -i \omega t) + b_{\omega,\vec k}^{*}(x) \exp (-i \vec{k}\cdot \vec{\varrho} + i \omega t)$,
where $\vec{r}=(x,\vec{\varrho})$, $\vec{\varrho}$ lies in the plane orthogonal to the $x$ direction (we consider the situation that all functions decay exponentially with increasingly larger positive $x$), $\vec{k}$ is the wave vector along this plane and * denotes complex conjugation. We will suppress the indices and simplify the notation by using $a$ and $b$ instead of $a_{\omega,\vec k}(x)$ and $b_{\omega,\vec k}(x)$. Introducing the
functions $\psi_1 = a + b$ and $\psi_2 = a - b$, after linearizing  Eq.~(\ref{eq:s0})  we get a
pair of coupled Schr\"{o}dinger equations:\cite{Pikhitsa}
\bea
-\frac{1}{2}\frac{d^{2}}{dx^{2}}\psi_{1}+(3\psi_{0}^{2}-1+\kappa^{2})\psi_{1} &=& \varepsilon \psi_{2} ,
\label{eq:s1} \\
-\frac{1}{2}\frac{d^{2}}{dx^{2}}\psi_{2}+(\psi_{0}^{2}-1+\kappa^{2})\psi_{2} &=& \varepsilon \psi_{1},
\label{eq:s2}
\ea
where $\kappa=|\vec {k}|\xi/\sqrt{2}=k\xi/\sqrt{2}$ and $\varepsilon=\omega$. This pair of equations is identical to the corresponding Bogoliubov-de Gennes equations (see \cite{Dziarmaga,Chen}) if one rewrites them for the functions $a$ and $b$.  To the best of our knowledge, Eqs.~(\ref{eq:s1})~and~(\ref{eq:s2}) have never been solved exactly before for arbitrary nonzero $\kappa$ and $\varepsilon$.
We find a formal general solution for these equations and illustrate its viability by obtaining a rigorous expression for the spectrum of localized phonons.

The spectrum of bulk excitations can be easily found from (\ref{eq:s1})~and~(\ref{eq:s2}) when neglecting the derivative terms far from the boundary $x=0$ to obtain the well-known Bogoliubov spectrum
$
\varepsilon_{\rm b}= \kappa \sqrt{2+\kappa^2}.
$
For $\kappa \to 0$ this gives the bulk phonon dispersion
$\varepsilon_{\rm b} \simeq  \sqrt{2}\kappa+\kappa^3/2\sqrt{2}$ and for $\kappa \to \infty$ 
it reads $\varepsilon_{\rm b} \simeq \kappa^2 +1$, which represents a free boson plus chemical potential.
The localized excitations to be derived, by definition, have an energy spectrum lying lower than the bulk one.

\subsection{Supersymmetry at an exceptional point}
We first remark that at the exceptional point of symmetry $\varepsilon=0$ and  $\kappa= 0$,
Eqs.~(\ref{eq:s1})~and~(\ref{eq:s2}) are the parts of a supersymmetric Hamiltonian with zero ground state energy. Indeed, on introducing the matrix operator
\bea
\hat{A}= \left( \begin{array}{cc}
-\frac{1}{\sqrt{2}}\frac{d}{dx} - \sqrt{2}\psi_{0} &  0\\
0 & -\frac{1}{\sqrt{2}}\frac{d}{dx}+ \frac{1}{\sqrt{2}}\frac{1-\psi_{0}^{2}}{\psi_{0}}  \end{array} \right) \label{eq:s2a}
\ea
so that the left-hand side of Eqs.~(\ref{eq:s1})~and~(\ref{eq:s2}) takes the form of a matrix Hamiltonian
\bea
\hat H_{-} =\hat{A}^{\dag} \hat{A}= \left( \begin{array}{cc}
-\frac{1}{2}\frac{d^{2}}{dx^{2}}+3\psi_{0}^{2}-1&  0\\
0 & -\frac{1}{2}\frac{d^{2}}{dx^{2}}+\psi_{0}^{2}-1  \end{array} \right)\nn
\label{eq:s2b}
\ea
with its partner Hamiltonian
\bea
\hat H_{+} =\hat{A} \hat{A}^{\dag}= \left( \begin{array}{cc}
-\frac{1}{2}\frac{d^{2}}{dx^{2}}+\psi_{0}^{2}+1&  0\\
0 & -\frac{1}{2}\frac{d^{2}}{dx^{2}}+\frac{1-\psi_{0}^{2}}{\psi_{0}^{2}}  \end{array} \right) , 
\label{eq:s2c}
\ea
we produce a supersymmetric (SUSY) Hamiltonian
\bea
\hat H_{\rm SUSY} =\left( \begin{array}{cc}
\hat H_{-}&  0\\
0 & \hat H_{+}  \end{array} \right)
\label{eq:s2d}
\ea
that may canonically be expressed through the supercharges
\bea
\hat Q = \left( \begin{array}{cc}
0 &  0\\
\hat{A} & 0 \end{array} \right),\,
\hat Q^{\dag} =\left( \begin{array}{cc}
0 &  \hat{A}^\dag\\
0 & 0 \end{array} \right)
\label{eq:s2e}
\ea
as an anticommutator
\bea
\hat H_{\rm SUSY}=\{Q,Q^\dag \};
\hat Q^2=0, (\hat Q^{\dag})^2=0 .
\label{eq:s2f}
\ea
The supersymmetry is explicitly broken when either $\varepsilon$ or  $\kappa$ (or both)
are not zero which, as we will discuss in detail below, leads to a splitting of the SUSY-degenerate  ground state into two gapless excitations (a ``light" one with $\varepsilon \propto \kappa $ and a ``heavy" one with $\varepsilon \propto \kappa^{3/2} $), both bound to the 
wall.\cite{Pikhitsa}

\subsection{Boundary conditions}
The boundary conditions for $\psi_1$ and $\psi_2$ in
\eqref{eq:s1} and \eqref{eq:s2} form two
distinct classes.   At the node of the kink $\psi=0$, that is both ${\rm Re}~ \psi=0$ and ${\rm Im}~ \psi=0$, and therefore also $\psi_1=0$ and $\psi_2=0$.

However, an additional possibility exists: For  $\varepsilon=0$ and $\kappa=0$, Eqs.~(\ref{eq:s1})~and~(\ref{eq:s2}) have the solutions $\psi_1^{(0)}=1-\psi_0^2$ and $\psi_2^{(0)}=\psi_0$, the first of which is the so-called ``zero mode",\cite{Pikhitsa,Dziarmaga}
which leads to Goldstone gapless modes (ripplons and phonons) when the SUSY
is broken.
This corresponds to a translation of the kink $\psi_0$ as a whole along $x$, resulting in the
displaced kink $\psi_0$ to read as follows: $\psi_0(x+\delta x)\simeq  \psi_0(x)+\psi_1^{(0)} \delta x$. Thus the condition ${\rm Re}~ \psi=0$ turns into $\psi_0' \delta x(\vec \varrho, t)+ {\rm Re}~ \vartheta (\vec r, t)=0$ which determines the shape of the loci of nodes $\delta x(\vec \varrho, t)$ (the shape of the surface). The derivative of such a mode with respect to $x$ is zero at $x=0$. The mode with the mixed boundary conditions
$\frac{d}{dx}\psi_1\mid_{x=0}=0 $ and $\psi_2\mid_{x=0}=0$
allows the ``rippling" of the soliton and is thus called {\em ripplon} mode.\cite{Pikhitsa}
As we shall see below, its energy spectrum at low $\kappa $ coincides with the one for
a classical capillary wave. The mode with ``zero"  boundary conditions $\psi_1\mid_{x=0}=0$ and $\psi_2\mid_{x=0}=0$, which correspond to a flat hard wall will be called {\em surface phonon} mode (with a spectrum starting linear).\cite{Pikhitsa}  Finally, the flat hard wall 
excludes the possible solution $x\psi_0 -1$ \cite{Muryshev}  of Eq.~(\ref{eq:s2}) at $\kappa=0$, $\varepsilon=0$ which could lead to the so-called snake instability,\footnote{The snake instability amounts to a {\em moving} wall (a nodal plane) with its transverse parts moving at different velocities, which is hence acting to destroy the wall} cf. Refs.~\cite{Kuznetsov,Muryshev}. 
This latter solution does not satisfy zero boundary conditions. 

\section{Asymptotic solutions} \label{approx}
We first derive the large and small wavelength solutions of the BdGE, noting that solely the large  wavelength case has been considered before.\cite{Pikhitsa,Nambu}
\subsection{Large wavelengths}
\begin{figure}[t]
 \vspace*{-2em}
 \includegraphics[width=0.5 \textwidth]{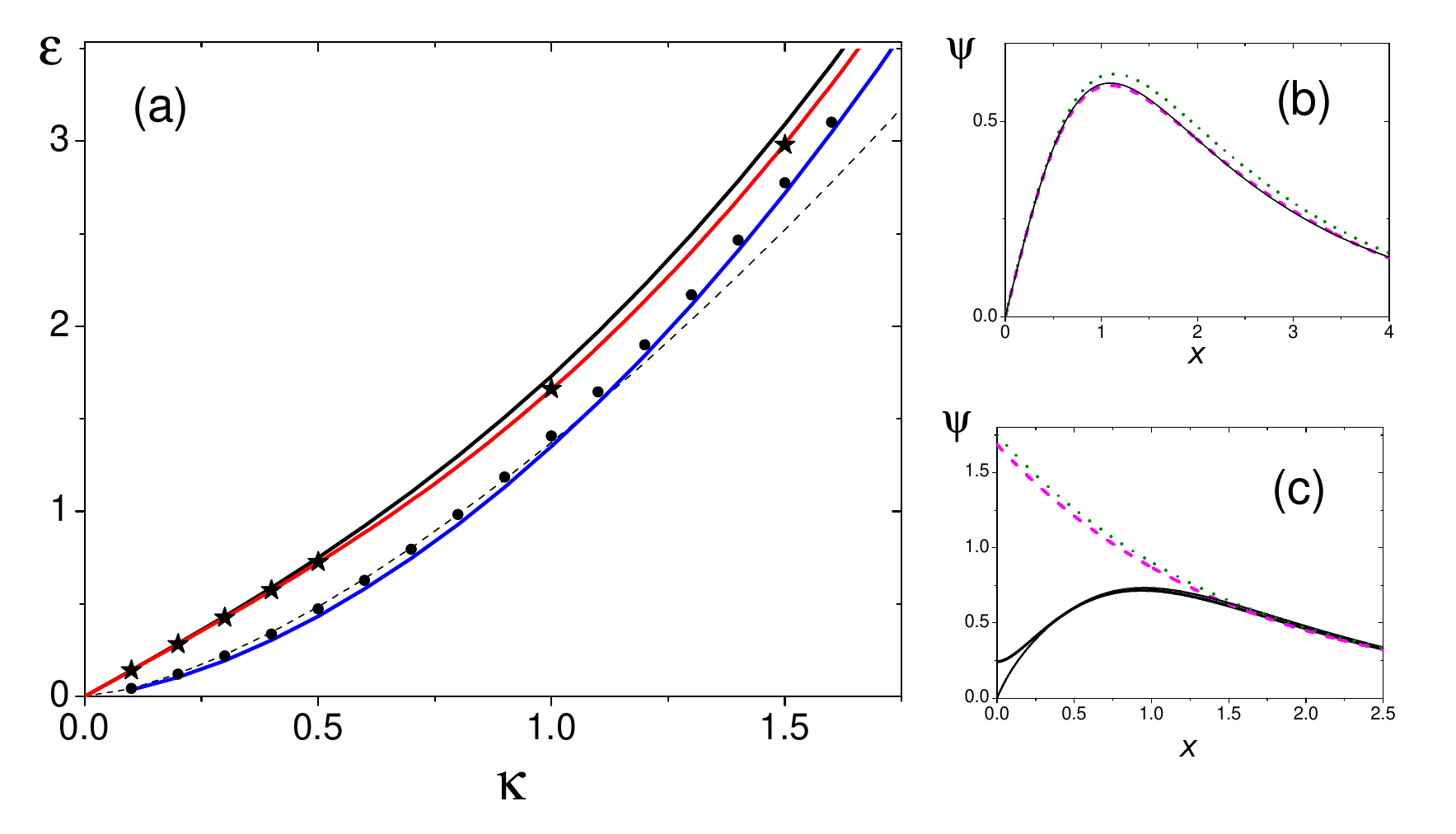}
  \caption{The small momentum part of the dimensionless spectra
of elementary excitations vs dimensionless wave number, together with 
surface phonon ($\equiv$ flat hard wall) and ripplon ($\equiv$ flexible wall) wave functions at larger momenta. 
  (a) The solid black line is the Bogoliubov bulk excitation spectrum and the {black} dashed line is the capillary wave spectrum (\ref{eq:rip}). The circles mark the  spectrum of the ripplon calculated by numerically solving Eqs.~(\ref{eq:s1})~and~(\ref{eq:s2}), and the blue line shows the first approximation represented 
  by Eqs.~\eqref{eq:s20} and \eqref{eq:s21}.
Finally, the stars mark the numerical spectrum of the surface phonon, and the red line is the exact solution (\ref{eq:s18}).
(b) The numerical wavefunctions of the surface phonon at $\kappa=3.5$ are shown with dashed ($\psi_1$) and dotted ($\psi_2$) lines together with $\psi_\infty$ (solid line) which they approach at large $x$. (c) The numerical wavefunctions of the ripplon mode at $\kappa=5$. It is seen that $\psi_1$ (thick solid line) lies very close to $\psi_2$ except the coordinate origin where $\psi_1$ has zero derivative. The dashed purple and dotted dark green lines show the asymptotic behavior $\propto\exp(-\alpha_2 x)$ for $\psi_{1,2}$.}\label{fig:rpl}
\end{figure}

First consider the case of $\kappa \to 0$.  For the ripplon spectrum we make an ansatz
for $\psi_{1,2}$ in the form of a series in $\varepsilon$:
$\psi_1\simeq  \psi_1^{(0)} + \varepsilon \psi_1^{(1)} + \mathcal{O}(\varepsilon^2)$ and
$\psi_2\simeq  \psi_2^{(0)} + \varepsilon \psi_2^{(1)} + \mathcal{O}(\varepsilon^2)$.
A zeroth-order approximation is the solution of the homogeneous equations Eqs.~(\ref{eq:s1})~and~(\ref{eq:s2}) with $\varepsilon=0$. This solution can be found for any $\kappa$ (which is verified by direct substitution): 
\bea
\label{eq:z1}
\psi_1^{(0)}=A \exp(-\alpha_1x)\left(\frac{\alpha_1^2-1}{3}+\alpha_1\psi_0 + \psi_0^2\right),\\
\label{eq:z2}
\psi_2^{(0)}=B \exp(-\alpha_2x)(\psi_0 + \alpha_2),
\ea
where $\alpha_1=\sqrt{2}\sqrt{2+\kappa^2}$ and $\alpha_2=\sqrt{2}\kappa$.
To determine $\psi_1^{(1)}$ and $\psi_2^{(1)}$, we have to solve the inhomogeneous equations that follow from Eqs.~(\ref{eq:s1}), (\ref{eq:s2}) when $\kappa=0$:
\bea
\label{eq:ss1}
-\frac{1}{2}\frac{d^{2}}{dx^{2}}\psi_1^{(1)}+\left(3\psi_{0}^{2}-1\right)\psi_{1}^{(1)} &=& B\psi_{0}, \\
 \label{eq:ss2}
-\frac{1}{2}\frac{d^{2}}{dx^{2}}\psi_{2}^{(1)}+\left(\psi_{0}^{2}-1\right)\psi_{2}^{(1)} &=&  A\left(1 - \psi_{0}^2\right).
\ea
With the help of the Green functions of the homogeneous equations the inhomogeneous solutions are found as:
\bea
\label{eq:az1}
\psi_1^{(1)}=\frac{1}{2}B\left\{\psi_0+x(1 - \psi_0^2)\right\},\\
\label{eq:az2}
\psi_2^{(1)}=-A.
\ea
Finally, the derivative with respect to $x$ of $\psi_1$ at $x=0$ is found from Eqs.~(\ref{eq:z1}), (\ref{eq:az1}) to be $\psi_1'=A\alpha_1(2-\alpha_1)(2+\alpha_1)/3 + B \varepsilon $, which according to the mixed boundary conditions should be zero together with $\psi_2=-A\varepsilon + B \alpha_2$, according to Eqs.~(\ref{eq:z2})~and~(\ref{eq:az2}). 
A vanishing determinant of the $A,\,B$ linear equations matrix 
\bea
\det \left( \begin{array}{cc}
\alpha_1(2-\alpha_1)(2+\alpha_1)/3 & \varepsilon\\
-\varepsilon & \alpha_2 \end{array} \right)=0
\label{eq:zz3}
\ea
gives the ripplon spectrum. Taking into account that $\alpha_1 \simeq  2 + \kappa^2/2$ for $\kappa\to 0$ and retaining only the lowest power of $\kappa$, we obtain the fractional dispersion
\bea
\varepsilon=\sqrt{\frac{4\sqrt{2}}{3}}\kappa^{3/2}.
\label{eq:rip}
\ea
The spectrum (\ref{eq:rip}) is shown in Fig.~\ref{fig:rpl}. Note that the localization of the ripplon at low $\kappa$ is governed by $\alpha_2=\sqrt{2}\kappa$.
The spectrum (\ref{eq:rip}) coincides with the well-known expression for the frequency of capillary waves (in the deep water limit),
which reads in dimensionful form 
$\varepsilon = \hbar \sqrt{\sigma/mn_0}~ k^{3/2}$ where $\sigma=\frac23\hbar c n_0$ is the surface energy density of the stationary soliton $\psi_0$. \cite{Ginzburg1} We note that $\sigma$ is exactly half of the energy of the dark soliton at rest [see Eq.~(5.59) in \cite{Pitaevskii1}].

Zero boundary conditions lead to surface phonons, for which we obtain the whole spectrum analytically in Sec.~\ref{phonon_Sec}  below. 
We here only mention in connection to the above discussion that $\alpha_2$ for phonons at low $\kappa$ is proportional to $\kappa^2$, indicating a much weaker localization as compared to the ripplons.

\subsection{Small wavelengths}
When $\kappa \rightarrow \infty$, we introduce the function $\chi$ and constant $\Delta$ so that $\psi_{1}=\psi_{2}+\chi/k^2$, $\psi_{2}=\psi_\infty$ and $\epsilon=\kappa\sqrt{\kappa^2+2}-\Delta \asymp \kappa^2+1-\Delta $. Then Eqs.~(\ref{eq:s1}) and (\ref{eq:s2}) turn into
 \bea
 \label{eq:s7}
-\frac{1}{2}\frac{d^{2}}{dx^{2}}\psi_{\infty}+(3\psi_{0}^{2}-2+\Delta)\psi_{\infty} &=& - \chi ,\\
\label{eq:s8}
-\frac{1}{2}\frac{d^{2}}{dx^{2}}\psi_{\infty}+(\psi_{0}^{2}-2+\Delta)\psi_{\infty} &=& \chi ,
\ea
with $\chi=-\psi_0^2 \psi_\infty$,
which after adding and subtracting both equations leads to
\begin{multline}
\psi_\infty = (1-\psi_0^2)^{\alpha_{\infty} /2} \\
\times {}_{2}F_{1}\left(\alpha_{\infty} - s,\alpha_{\infty}+s+1, \alpha_{\infty} + 1, \frac{1-\psi_0}{2} \right),
\label{eq:s10}
\end{multline}
where the hypergeometric function contains $\alpha_{\infty} = \sqrt{2\Delta}$ and $s=(\sqrt{17}-1)/2$ is one of the solutions of the equation $s(s+1)=4$ (see \cite{Landau}). The second solution leads to the same result.
The boundary condition $\psi_2=0$ at $x=0$ imposes the following identity:  
\begin{multline}
{}_{2}F_{1}\left(\alpha_{\infty} - s,\alpha_{\infty}+s+1, \alpha_{\infty} + 1, \frac{1}{2} \right) = \\ \frac{\Gamma(\frac{1}{2})\Gamma(\alpha_{\infty} +1)}{\Gamma(\frac{1}{2}[1+\alpha_{\infty}-s])\Gamma(\frac{1}{2}[2+\alpha_{\infty}+s])}= 0.
\label{eq:s11}
\end{multline}
which demands (for fixed $s$)  $1+\alpha_{\infty} -s=0$ in order to have the infinity in the denominator from the corresponfing Gamma function, and therefore $\alpha_{\infty}=(\sqrt{17}-3)/{2}\simeq  0.562 $ while $\Delta=\alpha_\infty^2/2\simeq  0.158 $. Finally, the hypergeometric function in (\ref{eq:s10}) reduces to $\psi_0$ so that $\psi_\infty=\psi_0(1-\psi_0^2)^{\alpha_{\infty}/2}=\tanh(x)/\cosh(x)^{\alpha_{\infty}}$ (see Fig.\ref{fig:rpl} (b)) and $\chi=-\psi_0^3(1-\psi_0^2)^{\alpha_{\infty}/2}$. Therefore, both $\psi_2$ and $\psi_1$ satisfy the zero boundary conditions. Analogously, one can show\footnote{By making use of the known solutions of the homogeneous equations (\ref{eq:z1})~and~(\ref{eq:z2}) to satisfy the boundary conditions} that the function $\psi_\infty$ is also the limiting function for large $\kappa$ in the case of mixed boundary conditions, so that the difference between the functions appears only in close proximity to the boundary $x=0$,
at a typical distance $1/\kappa$ [see Fig.\ref{fig:rpl}\,(c);
 $\psi_1$ deviates from $\psi_2$ and hits the $\psi$ axis with zero derivative].
Thus  the dimensionful binding energy of the  excitation localized near the surface depends only on the bulk parameter $mc^2$. 

\section{Exact solution of the full BdGE}
{It is well established that many exact solutions of 
Schr\"odinger equations with various types of potentials can be directly related to solutions of hypergeometric equations (see, e.g., Ref.\cite{Ishkhanyan_2015} for a list); hence factorizations used in quantum mechanics can be obtained from factorizations employing hypergeometric operators. \cite{Cotfas} Here, using {\em hypergeometric matrices} (which we discuss in detail in the Appendix), we derive below an exact solution of the BdGE.} 

We aim at finding the exact solution of Eqs.~(\ref{eq:s1}) and (\ref{eq:s2}) at {\em arbitrary} nondimensionalized momentum  $\kappa$. To do so, let us transform these equations into a single matrix hypergeometric equation, where we employ the fact that matrix generalizations of both hypergeometric function and Gamma function were previously shown to be mathematically viable tools.\cite{Tirao,Jodar}
We introduce {a wavefunction ansatz by analogy with Eq.~(\ref{eq:s10}): }
\bea
\psi_{1,2}=(1-\psi_0^2)^{\alpha/2} \phi_{1,2}, qquad
z=\frac{1-\psi_0}2, 
\ea
{with a formal parameter $\alpha$. Below this single scalar parameter will be replaced with a matrix, which constitutes the key starting point of finding our exact solution.} 
We now rewrite Eqs.~\eqref{eq:s1} and \eqref{eq:s2} as 
\begin{multline}
z(1-z)\frac{d^2\phi_1}{dz^2} + \left\{\alpha+1 - 2(\alpha+1)z\right\}\frac{d\phi_1}{dz}\\
+\left\{ 6 - \alpha(\alpha + 1)\right\}\phi_1
+\frac{1}{2z(1-z)}\left(\frac{\alpha^2}{2} - 2- {\kappa^2}\right)\phi_1\\
={\varepsilon}\frac{\phi_2}{2z(1-z)},\\
z(1-z)\frac{d^2\phi_2}{dz^2} + \left\{\alpha + 1 - 2(\alpha+1)z\right\}\frac{d\phi_2}{dz}\\
+ \left\{2 - \alpha(\alpha + 1)\right\}\phi_2
+\frac{1}{2z(1-z)}\left(\frac{\alpha^2}{2} - {\kappa^2}\right)\phi_2\\
={\varepsilon}\frac{\phi_1}{2z(1-z)}.
\label{eq:s12}
\end{multline}
To turn (\ref{eq:s12}) into a matrix hypergeometric equation, we introduce the vector-function $\hat{\Phi}$, the identity matrix $\hat {1}$, the matrix $\hat\alpha$, and matrices $\hat a,\hat b, \hat c$ derived from it, as follows
\bea
\hat {\Phi} =  {\left( \begin{array}{c}
\phi_1 \\
\phi_2 \end{array} \right)} ,\\
\hat {\alpha}^2= 2\left( \begin{array}{cc}
2+{\kappa^2} &  {\varepsilon}\\
{\varepsilon} & {\kappa^2} \end{array} \right),\\
\hat {c}=\hat{\alpha}+\hat{1},\\
 \label{eq:sa13}
 \hat{1}+\hat{a}+\hat{b}=2(\hat{\alpha}+\hat{1}),\\
 -\hat{a}\hat{b}= \left( \begin{array}{cc}
6 &  0\\
0 & 2 \end{array} \right)- \hat{\alpha}^2-\hat{\alpha}.
\label{eq:s13}
\ea
Taking the square root of the matrix $\hat{\alpha}^2$ gives, choosing the positive sign,
\bea
\hat{\alpha} 
= {2}\left( \begin{array}{cc}
r &  l\\
l & p \end{array} \right),
\label{eq:s13b}
\ea
where $r=\sqrt{1+\kappa^2/2-l^2}$, $ p= \sqrt{\kappa^2/2-l^2}$, and $ l={\varepsilon}\sqrt{\kappa^2+1-(\kappa^2(\kappa^2+2)-\varepsilon^2)^{1/2}}/2\sqrt{\varepsilon^2+1}$. 
The two positive eigenvalues of the matrix $\hat \alpha$ are
\bea
 \alpha_{1,2} = \sqrt{2}\sqrt{1+\kappa^2 \pm \sqrt{\varepsilon^2+1}}
\label{eq:sa14}
\ea
and $\exp (-\alpha_2 x)$ determines the asymptotic decay  of $\psi_1$ and $\psi_2$ as $x\to \infty$ 
(corresponding to the lower sign above). 
After introducing the matrices, Eq.~(\ref{eq:s12}) {becomes the canonical  Gauss hypergeometric equation in matrix form}
\bea
z(1-z)\hat{\Phi}''+(\hat{c}-(\hat{1}+\hat{a}+\hat{b})z)\hat{\Phi}' -\hat{a}\hat{b}\hat{\Phi}=0,
\label{eq:s14}
\ea
where primes mean differentiation with respect to $z$.

\begin{figure}[b]
\includegraphics[width=0.5\textwidth]{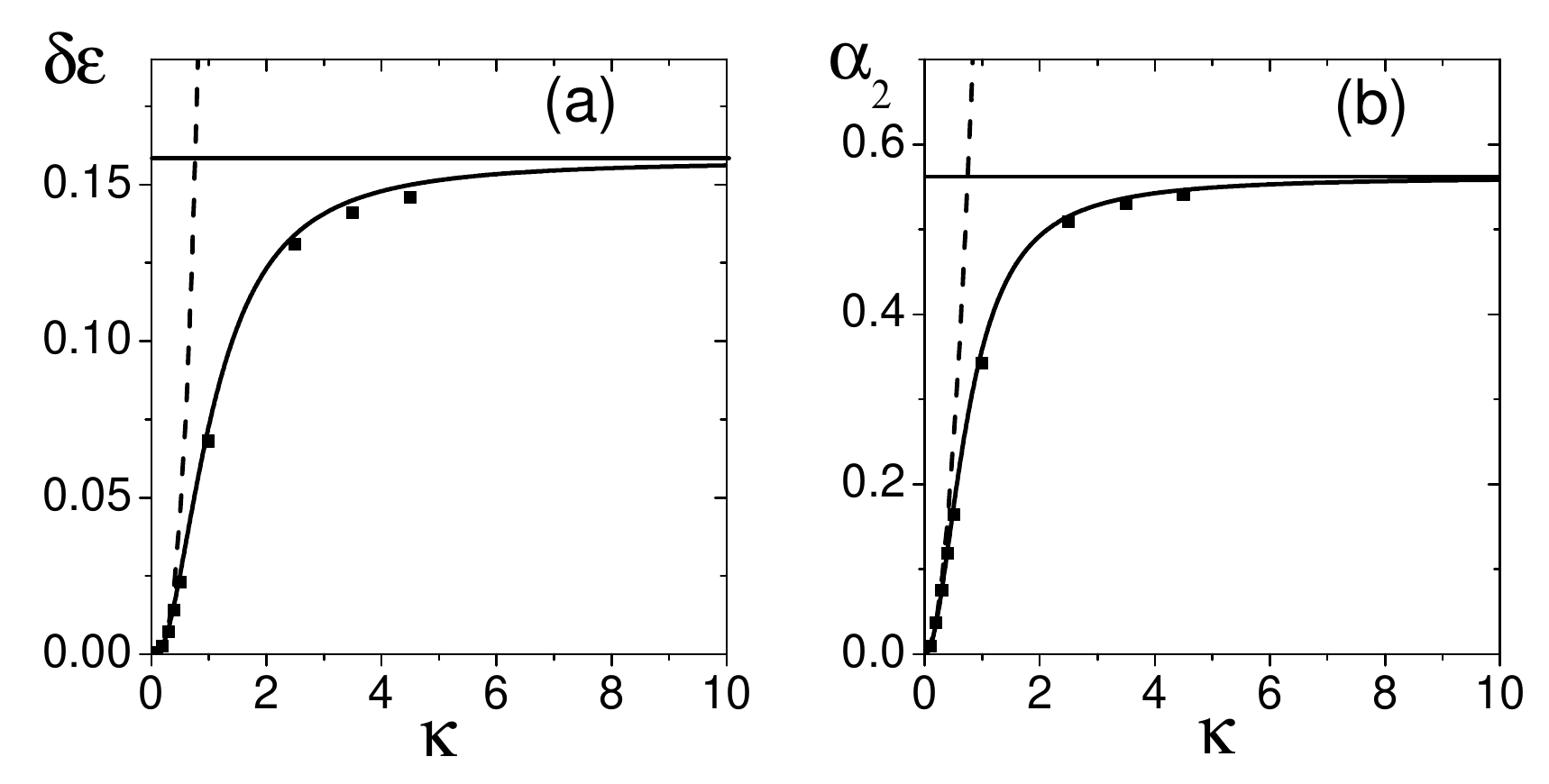}
\caption{
{\em Surface phonon.} (a)~Binding energy 
vs wavenumber. The solid line is the result of the exact dispersion given by Eq.~(\ref{eq:s18}). The solid squares show the results of the numerical solution of Eqs.~\eqref{eq:s1}
and \eqref{eq:s2} with zero boundary conditions.  The horizontal line is at $\Delta$; the dashed line is  $\kappa^3/2\sqrt{2}$ from the small $\kappa$ behavior. 
(b)~ Solid line: Exact solution for $\alpha_2$ as given by Eqs.~\eqref{eq:sa14} and \eqref{eq:s18}. The horizontal line is at $\alpha_\infty$; the dashed line is $\kappa^2$.}
\label{fig:phonenerg}
\end{figure}
\subsection{Surface phonons}\label{phonon_Sec} 
{Eq.~(\ref{eq:s14}) is formally  solved by Eq.~\eqref{eq:saa14} contained in the Appendix.} 
We can then obtain the spectrum of the surface phonon localized
near a flat hard wall as follows.  
The boundary condition at $x=0$ (that is at $z=1/2$) will be fulfilled when $\hat \Phi=0$, {which demands that the determinant of matrix function (\ref{eq:s16}) be zero. 
The spectrum is then given by the equation} 
\bea
\det \hat{P} =(3r+\kappa^2)(3p + \kappa^2)-(3l +\varepsilon)^2 =0,
\label{eq:s18}
\ea
{where the matrix $\hat{P}$ is derived in the Appendix, see Eq.~\eqref{eq:s17}.}
Eq.~(\ref{eq:s18}) reproduces the spectrum calculated before for the two limiting cases
$\kappa \to 0$ and $\kappa \to \infty$ in section~\ref{approx}.
When $\kappa \to 0$ the spectrum is $\varepsilon = \sqrt{2}\kappa + \mathcal{O}(\kappa^5)$, so that the $\kappa^3$ term is missing,
while the bulk phonon starts with higher energy as $\varepsilon_{\rm b} = \sqrt{2}\kappa + \kappa^3/2\sqrt{2} + \mathcal{O}(\kappa^5)$. Let us define the binding energy as $\delta\varepsilon = \varepsilon_{\rm b}-\varepsilon$. Then the latter starts as $\kappa^3/2\sqrt{2}$ [see Fig.~\ref{fig:phonenerg}\,(c)]. Now let us consider the other limit $\kappa \to \infty$. It is easy to see that seeking the solution in the form $\varepsilon \asymp \kappa^2+1-\Delta$  leads to $r=p \asymp \kappa/2 + \sqrt{2\Delta}/4$ and $l\asymp \kappa/2 - \sqrt{2\Delta}/4$ which after substitution into Eq.~(\ref{eq:s18}) give $ 2\Delta+3\sqrt{2\Delta}-2=0.$
This has the same root as found before from Eq.~(\ref{eq:s11}), $\sqrt{2\Delta}=(\sqrt{17}-3)/2=\alpha_{\infty}\simeq  0.562$ and therefore $\delta\varepsilon_{\infty}=\Delta=\alpha_{\infty}^2/2\simeq  0.158$.

The coincidence with the exact asymptotic results obtained in section~\ref{approx} confirms the correctness of Eq.~(\ref{eq:s18}).
Note that the slower decay exponent $\alpha_2$ can be approximated by a simple expression
$\alpha_2=\kappa^2/(1+\kappa^2/\alpha_\infty)$ that fits the exact expression of Eq.~(\ref{eq:sa14}) with $\varepsilon$ from the exact solution of Eq.~(\ref{eq:s18}) within $0.2 \%$. We plot the decay exponential in Fig.~\ref{fig:phonenerg}(d) {in a broad range of wavenumbers}.

Finally, it is interesting to note that the first order approximation in powers of $z=1/2$
in the case of the flat hard wall boundary conditions for surface phonons can be represented by the equation
\begin{multline}
\hspace*{-1em} \left(\hat{1}+\frac{\hat{U_1}}2\right)_{0,0} \left(\hat{1}+\frac{\hat{U_1}}2\right)_{1,1}-
  \left(\hat{1}+\frac{\hat{U_1}}2\right)_{0,1} \left(\hat{1}+\frac{\hat{U_1}}2\right)_{1,0} \\=0 
\label{eq:s22}
\end{multline}
which, distinct from the case of ripplons discussed below, {\em accidentally} gives the exact spectrum of Eq.~\eqref{eq:s18}, where $\hat U_1$ is defined below in Eq.~\eqref{eq:s21}. 

\begin{figure}[t]
\includegraphics[width=0.5\textwidth]{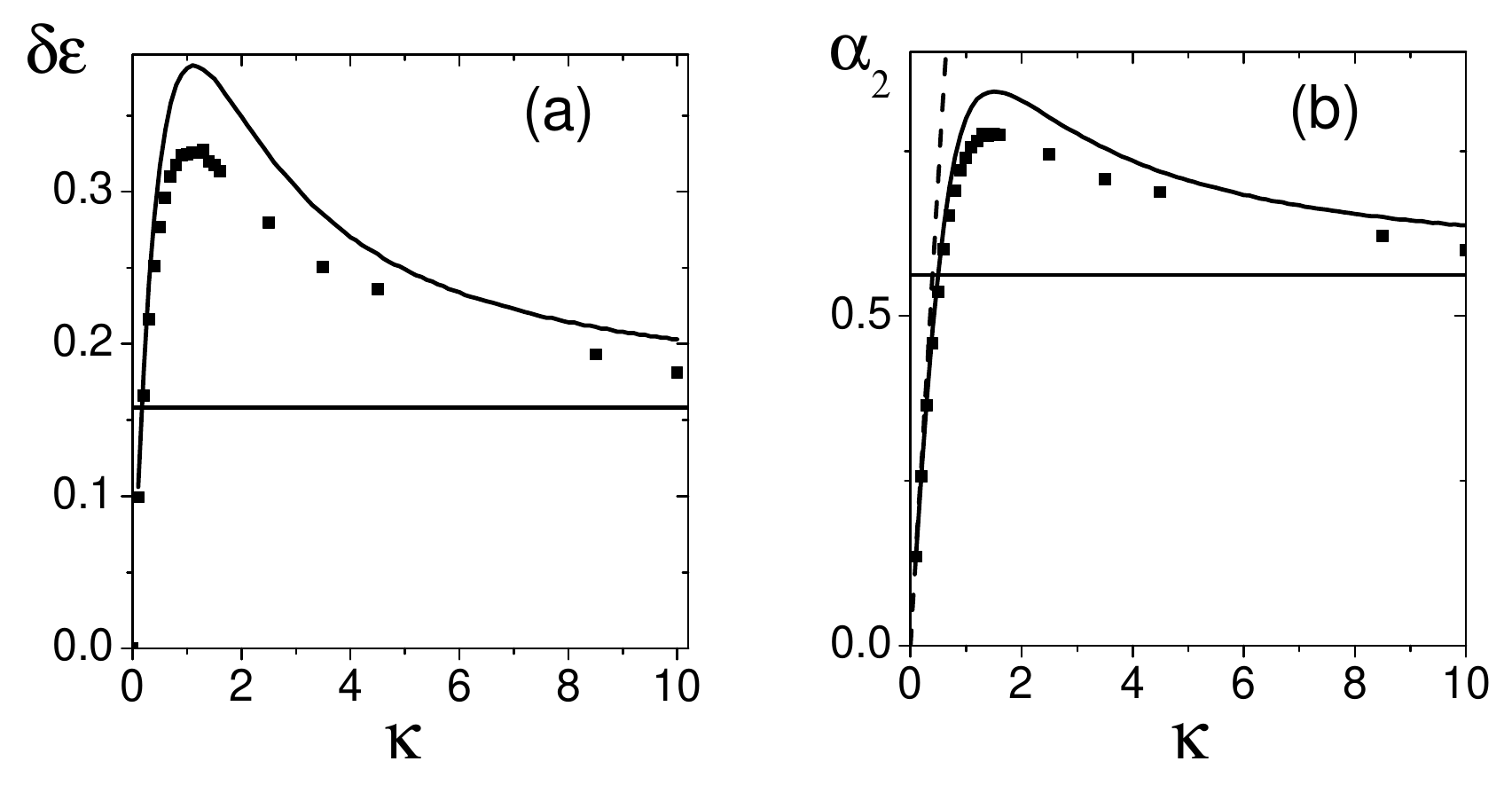}
\caption{{\em Ripplon.} (a)~Dimensionless binding energy 
vs the dimensionless wavenumber.
 The solid squares show the results of the numerical solution of Eqs.~\eqref{eq:s1} and \eqref{eq:s2} with mixed boundary conditions. The solid line is the result of using the approximation provided by Eqs.~\eqref{eq:s20} and \eqref{eq:s21}. 
  The horizontal line is at energy $\Delta = 0.158$. (b) The squares show the decay parameter $\alpha_2$ 
 obtained with  Eq.~(\ref{eq:sa14}). The solid line is again the result of the first approximation of the exact solution given by Eq.~(\ref{eq:s21}). The horizontal line is at $\alpha_\infty = 0.568$. The analytical relation $\alpha \simeq  \sqrt{2}\kappa$ at small $\kappa$ is shown by the dashed line.}
\label{fig:rippenerg}
\end{figure}
\subsection{Ripplons}
Now consider the case of mixed boundary conditions corresponding to ripplons. Let us first rephrase the general form of the solution  \eqref{eq:saa14} provided in the Appendix 
in the form of a vector function.
With the help of (\ref{eq:s16}), we get
 \bea
{\left( \begin{array}{c}
\phi_1 \\
\phi_2 \end{array} \right)} = {}_2F_1(a,b,c,z) \,{\left( \begin{array}{c}
A \\
B \end{array} \right)} , \nn
\label{eq:s18a}
\ea
where $A$ and $B$ are arbitrary constants.

Writing the surface phonon boundary conditions explicitly, 
\bea
\phi_1 &=& A[{}_2F_1(a,b,c,z)]_{0,0}+B[{}_2F_1(a,b,c,z)]_{0,1} =0,\nn
\phi_2 &=& A[{}_2F_1(a,b,c,z)]_{1,0}+B[{}_2F_1(a,b,c,z)]_{1,1} =0,\nn
\label{eq:s19}
\ea
the spectrum for surface phonons is obtained after equating
the determinant of the above equation for $A,\,B$ to zero at
$z = 1/2$.
We then proceed analogously as for this case of the
flat wall for the rippled wall, except that we differentiate with
 respect to  $z$ the first equation for $\phi_1$.

Expanding the hypergeometric function of Eq.~\eqref{eq:saa14} and its $z$ derivative to first order in $z=1/2$, one gets from Eq.~(\ref{eq:s19}) 
\begin{multline}
\hspace*{-1em} \left(\hat{U_1}+\frac{\hat{U_2}}2\right)_{\!0,0} \!
 \left(\hat{1}+\frac{\hat{U_1}}2\right)_{\!1,1}-
 \left(\hat{U_1}+\frac{\hat{U_2}}2\right)_{\!0,1}\!
 \left(\hat{1}+\frac{\hat{U_1}}2\right)_{\!1,0} \\=0,
\label{eq:s20}
\end{multline}
where $\hat{U_1}$ and $\hat{U_2}$ are calculated according to Eq.~\eqref{eq:s15} for $m=0,1$ with $\hat{\alpha}$ and $\hat{a}\hat{b}$ taken from Eqs.~\eqref{eq:s13b} and \eqref{eq:s13}, respectively:
 \bea
\begin{array}{c}
\hat{U_2}=(\hat{\alpha}+2\hat{1})^{-1}(\hat{a}\hat{b}+2(\hat{\alpha}+\hat{1}))U_1\\
\hat{U_1}=(\hat{\alpha}+\hat{1})^{-1}\hat{a}\hat{b}.\end{array}
\label{eq:s21}
\ea

\subsection{Comparison to numerics}
The energy spectrum and binding energy--decay parameter for ripplons {in a broad range of wavenumbers} are shown in 
Figs.~\ref{fig:rippenerg}\,(a),(b), respectively, in comparison to their
values obtained  with numerical solutions of the differential equations Eqs.~(\ref{eq:s1})~and~(\ref{eq:s2}), represented by the symbols.
For the numerics, we used 
PTC's MathCad~11, applying proper boundary conditions at the surface, {\em imposing} an exponentially fast decay at  infinity (the latter leads to underestimate the binding energy values, see for a discussion below).
One can see that even to lowest nontrivial order in the series on $z=1/2$ [see Eq.~\eqref{eq:s20}],  the results shown by solid lines in Fig.~\ref{fig:rippenerg}\,(a),(b)
are rather close to the numerical solutions.

On the other hand, for the surface phonon,
the numerical 
results can be rendered closer
to the exact spectrum from \eqref{eq:s18}, as displayed in Figs.~\ref{fig:phonenerg}\,(a),(b), although a slight systematic deviation is still noticeable.
These deviations stem from the fact that the numerical solution of the differential equations
relies on the criterium of localization: the solution should decay into the bulk, implying that
another boundary condition is that the wavefunction should approach zero at infinity.
Numerical calculations are imposing boundary conditions at a finite distance, however large.
The numerics therefore slightly exaggerates the decay; hence the numerical energy is slightly lower than the exact energy at a given wavenumber.

\section{Conclusion}
{In summary, starting from the matrix hypergeometric equation (\ref{eq:s14}), we obtained its formally  exact solution at the boundary (\ref{eq:s16}). Many exactly solvable Schr\"{o}dinger equations with various potentials have solutions of the  hypergeometric variety. Often there are also supersymmetric partners in the Hamiltonian operator, as in the case of a hydrogen atom with its Coulomb potential. When a continuous wavenumber is present, creating a bandgap structure, gapless states that stem from (or are accompanied by) Goldstone zero energy modes may exist.\cite{Hasan}  
In the case of the BdGE that we considered here, 
we were not only able to obtain an exact solution, but also to express the dispersion relation of the Bogoliubov surface excitations for surface phonons in the  closed form of the algebraic equation (\ref{eq:s18}). We have furthermore shown that for ripplons, even a lowest nontrivial order truncation of the hypergeometric series produces results close to numerical solutions of the BdGE.}

We now discuss the relation of the analytically obtained binding energy of surface phonons
to the experimental finding of Ref.\cite{Prisk} for helium II confined by the hard walls of  cylindrical pores.
Even though the present BEC model with contact interactions
does not reproduce the roton minimum in the bulk dispersion curve,
it provides a correct estimate for the binding energy in the low-density surface region.
Indeed, the binding energy (the difference between bulk and surface excitation energies) has been measured to be $0.15$ meV 
at the roton-region wavevector 
$k=2/{\angstrom}$.\cite{Prisk}  The latter corresponds to $\kappa\simeq 1$, using the estimate $\xi=\hbar/mc \simeq 0.7 \angstrom$, with the ``bulklike" speed of sound $c=228\,$m/s at full pore.\cite{Prisk}  We can read off Fig.~2(a) a dimensionless binding energy of approximately 0.07  at $\kappa \simeq 1$, which agrees to good accuracy with the experimental value (using that $mc^2\simeq 2.2$\,meV). 
This agreement was obtained at small wavelengths, to which previous approaches did not apply, and which in the bulk correspond to the roton minimum. Therefore, while the latter bulk dispersion feature is not accurately described by our mean-field model, we conclude that the quantum mechanism of trapping excitations close to a surface gives the correct magnitude of the binding energy. We note that the
quantum mechanism of binding surface excitations occurs in the solid-state physics of electrons as well, where the bound states are called Tamm and Shockley states.\cite{Tamm,Shockley}

The present method for exactly solving the Bogoliubov-de Gennes equations
is potentially also useful in more sophisticated cases 
than the presently considered one. 
Further extensions of the present approach are for example conceivable by incorporating effectively nonlocal interactions modelling rotons, which occur in dilute quantum gases dominated by dipole-dipole interactions.\cite{Fischer}  
{Furthermore, it would be of 
interest to investigate to which extent the present matrix hypergeometric equation approach can be applied to other physical systems of current widespread interest. For instance, to topological insulators, superconductors, and even to exotic topological mechanical materials.\cite{Kane}} 

\acknowledgments
The work of PVP was supported by the Global Frontier Center for Multiscale Energy Systems funded by the National Research Foundation of Korea (NRF) 
Grant No.~2012M3A6A7054855. URF has been supported by the NRF under
Grant No.~2017R1A2A2A05001422.

\appendix*

\section{The matrix-valued hypergeometric function}

Equation (\ref{eq:s14}), {taking the canonical form of a hypergeometric equation}, has a formal solution as a matrix-valued hypergeometric function of Gauss \cite{Tirao}
\bea
{}_2F_1(a,b,c,z)=\sum_{n\geq 0} \frac{z^n}{n!}(\hat{a},\hat{b},\hat{c})_n ,
\label{eq:saa14}
\ea
where $(\hat{a},\hat{b},\hat{c})_{0}=\hat{1}$, and higher matrix coefficients are
\begin{multline}
(\hat{a},\hat{b},\hat{c})_{m+1} =
(\hat{c}+m\hat{1})^{-1}(\hat{a}+m\hat{1})(\hat{b}+m\hat{1})\\
\times (\hat{c}+m\hat{1}-\hat{1})^{-1}(\hat{a}+m\hat{1}-\hat{1})(\hat{b}+m\hat{1}-\hat{1})\ldots \hat{c}^{-1}\hat{a}\hat{b}.
\label{eq:s15}
\end{multline}
{In the above equation, the matrices are ordered in a specific way, taking into account their generally noncommutative  nature. The related intricate mathematical questions have been discussed in detail when introducing the matrix hypergeometric function in \cite{Tirao}. Yet this noncommutative nature is not important for our purpose of obtaining the dispersion relations,   inasmuch  as we deal to this end with the determinant of the matrix hypergeometric function. The latter determinant is expressed 
below through products of the matrix Euler Gamma function and its inverse in Eq.~(\ref{eq:s16}). 
The Euler Gamma function itself is in turn a product of matrices and their inverse according to Eq.~(A4). 
The determinant of the matrix hypergeometric function is  hence independent of the order of the matrices occurring in it.}

The hypergeometric function at $z=1/2$ can be expressed through the matrix Gamma
function\cite{Jodar}  [because $\hat {c}=(\hat{1}+\hat{a}+\hat{b})/2$, see Eq.(\ref{eq:sa13})], so that
\bea
{}_{2}F_{1}\left(\hat{a},\hat{b}, \hat{c}, \frac{1}{2} \right) = \frac{\Gamma(\frac{1}{2})\Gamma(\hat{c})}{\Gamma\left(\frac{1}{2}[\hat{1}+\hat{a}]\right) \Gamma\left(\frac{1}{2}[\hat{1}+\hat{b}]\right)}.
\label{eq:s16}
\ea
Then the condition of $\hat\Phi =0$ 
imposed for surface phonons in Sec.~\ref{phonon_Sec} implies that the matrix (\ref{eq:s16}) 
has an eigenvalue zero and that therefore its determinant vanishes. 
 We then use that matrix Gamma functions can be represented as follows  \cite{Jodar}
\bea
\Gamma(\hat M)=\lim_{n\to \infty} (n-1)! n^{\hat M}[\hat M (\hat M+\hat 1)\dots (\hat M+n \hat 1) ]^{-1} .
\nn \label{M}
\ea
The role of the matrix $\hat M$ is played by either $\hat 1 +\hat a$ or $\hat 1 +\hat b$. Because the determinant of (\ref{eq:s16}) is required to vanish, the determinant of a Gamma function in the denominator should be infinite.  By \eqref{M}, this is only possible if either the determinant
of $\hat 1 +\hat a$ or that of $\hat 1 +\hat b$ is {zero}.

One can prove that the determinant of either $\hat 1 +\hat a$ or $\hat 1 +\hat b$  being zero 
gives the same spectrum. However an analytical solution for the matrix equations (\ref{eq:sa13}), (\ref{eq:s13}) for $\hat a$ and $\hat b$ is difficult. To obtain analytical results we instead utilize the product $\hat{P}=((\hat{1}+\hat{a})(\hat{1}+\hat{b}))/2=(\hat{1}+\hat{a}+\hat{b}+\hat{a}\hat{b})/2$.
We readily get the matrix $\hat{P}$ from Eqs.~(\ref{eq:sa13}),(\ref{eq:s13})~and~(\ref{eq:s13b}):
\bea
\hat{P}= \left( \begin{array}{cc}
3r+\kappa^2 &  3l +\varepsilon\\
3l+\varepsilon & 3p + \kappa^2 \end{array} \right), 
\label{eq:s17}
\ea
and taking $\det \hat{P}=0$ yields Eq.~\eqref{eq:s18} of the main text.

\bibliography{ripplon12}

\begin{thebibliography}{37}%
\makeatletter
\providecommand \@ifxundefined [1]{%
 \@ifx{#1\undefined}
}%
\providecommand \@ifnum [1]{%
 \ifnum #1\expandafter \@firstoftwo
 \else \expandafter \@secondoftwo
 \fi
}%
\providecommand \@ifx [1]{%
 \ifx #1\expandafter \@firstoftwo
 \else \expandafter \@secondoftwo
 \fi
}%
\providecommand \natexlab [1]{#1}%
\providecommand \enquote  [1]{``#1''}%
\providecommand \bibnamefont  [1]{#1}%
\providecommand \bibfnamefont [1]{#1}%
\providecommand \citenamefont [1]{#1}%
\providecommand \href@noop [0]{\@secondoftwo}%
\providecommand \href [0]{\begingroup \@sanitize@url \@href}%
\providecommand \@href[1]{\@@startlink{#1}\@@href}%
\providecommand \@@href[1]{\endgroup#1\@@endlink}%
\providecommand \@sanitize@url [0]{\catcode `\\12\catcode `\$12\catcode
  `\&12\catcode `\#12\catcode `\^12\catcode `\_12\catcode `\%12\relax}%
\providecommand \@@startlink[1]{}%
\providecommand \@@endlink[0]{}%
\providecommand \url  [0]{\begingroup\@sanitize@url \@url }%
\providecommand \@url [1]{\endgroup\@href {#1}{\urlprefix }}%
\providecommand \urlprefix  [0]{URL }%
\providecommand \Eprint [0]{\href }%
\providecommand \doibase [0]{http://dx.doi.org/}%
\providecommand \selectlanguage [0]{\@gobble}%
\providecommand \bibinfo  [0]{\@secondoftwo}%
\providecommand \bibfield  [0]{\@secondoftwo}%
\providecommand \translation [1]{[#1]}%
\providecommand \BibitemOpen [0]{}%
\providecommand \bibitemStop [0]{}%
\providecommand \bibitemNoStop [0]{.\EOS\space}%
\providecommand \EOS [0]{\spacefactor3000\relax}%
\providecommand \BibitemShut  [1]{\csname bibitem#1\endcsname}%
\let\auto@bib@innerbib\@empty
\bibitem [{\citenamefont {Gross}(1961)}]{Gross1961}%
  \BibitemOpen
  \bibfield  {author} {\bibinfo {author} {\bibfnamefont {E.~P.}\ \bibnamefont
  {Gross}},\ }\bibfield  {title} {\enquote {\bibinfo {title} {{Structure of a
  quantized vortex in boson systems}},}\ }\href {\doibase 10.1007/BF02731494}
  {\bibfield  {journal} {\bibinfo  {journal} {Il Nuovo Cimento (1955-1965)}\
  }\textbf {\bibinfo {volume} {20}},\ \bibinfo {pages} {454} (\bibinfo {year}
  {1961})}\BibitemShut {NoStop}%
\bibitem [{\citenamefont {Pitaevskii}(1961)}]{Pitaevskii}%
  \BibitemOpen
  \bibfield  {author} {\bibinfo {author} {\bibfnamefont {L.~P.}\ \bibnamefont
  {Pitaevskii}},\ }\bibfield  {title} {\enquote {\bibinfo {title} {{Vortex
  lines in an imperfect Bose gas}},}\ }\href {http://dx.doi.org/1} {\bibfield
  {journal} {\bibinfo  {journal} {Sov. Phys. JETP}\ }\textbf {\bibinfo {volume}
  {13}},\ \bibinfo {pages} {451} (\bibinfo {year} {1961})}\BibitemShut
  {NoStop}%
\bibitem [{\citenamefont {Pitaevskii}\ and\ \citenamefont
  {Stringari}(2003)}]{Pitaevskii1}%
  \BibitemOpen
  \bibfield  {author} {\bibinfo {author} {\bibfnamefont {L.}~\bibnamefont
  {Pitaevskii}}\ and\ \bibinfo {author} {\bibfnamefont {S.}~\bibnamefont
  {Stringari}},\ }\href@noop {} {\emph {\bibinfo {title} {{Bose Einstein
  Condensation}}}}\ (\bibinfo  {publisher} {Oxford Univ. Press, NY},\ \bibinfo
  {address} {New York},\ \bibinfo {year} {2003})\BibitemShut {NoStop}%
\bibitem [{\citenamefont {Carusotto}\ and\ \citenamefont
  {Ciuti}(2013)}]{Carusotto}%
  \BibitemOpen
  \bibfield  {author} {\bibinfo {author} {\bibfnamefont {Iacopo}\ \bibnamefont
  {Carusotto}}\ and\ \bibinfo {author} {\bibfnamefont {Cristiano}\ \bibnamefont
  {Ciuti}},\ }\bibfield  {title} {\enquote {\bibinfo {title} {Quantum fluids of
  light},}\ }\href {\doibase 10.1103/RevModPhys.85.299} {\bibfield  {journal}
  {\bibinfo  {journal} {Rev. Mod. Phys.}\ }\textbf {\bibinfo {volume} {85}},\
  \bibinfo {pages} {299} (\bibinfo {year} {2013})}\BibitemShut {NoStop}%
\bibitem [{\citenamefont {Zakharov}(1968)}]{Zakharov}%
  \BibitemOpen
  \bibfield  {author} {\bibinfo {author} {\bibfnamefont {V.~E.}\ \bibnamefont
  {Zakharov}},\ }\bibfield  {title} {\enquote {\bibinfo {title} {Stability of
  periodic waves of finite amplitude on the surface of a deep fluid},}\ }\href
  {\doibase 10.1007/BF00913182} {\bibfield  {journal} {\bibinfo  {journal}
  {Journal of Applied Mechanics and Technical Physics}\ }\textbf {\bibinfo
  {volume} {9}},\ \bibinfo {pages} {190} (\bibinfo {year} {1968})}\BibitemShut
  {NoStop}%
\bibitem [{\citenamefont {Bogoliubov}(1947)}]{Bogoliubov}%
  \BibitemOpen
  \bibfield  {author} {\bibinfo {author} {\bibfnamefont {N.~N.}\ \bibnamefont
  {Bogoliubov}},\ }\bibfield  {title} {\enquote {\bibinfo {title} {{On the
  theory of superfluidity}},}\ }\href@noop {} {\bibfield  {journal} {\bibinfo
  {journal} {J. Phys.(USSR)}\ }\textbf {\bibinfo {volume} {11}},\ \bibinfo
  {pages} {23} (\bibinfo {year} {1947})}\BibitemShut {NoStop}%
\bibitem [{\citenamefont {Bogoliubov}(1958)}]{Bogoliubov1958}%
  \BibitemOpen
  \bibfield  {author} {\bibinfo {author} {\bibfnamefont {N.~N.}\ \bibnamefont
  {Bogoliubov}},\ }\bibfield  {title} {\enquote {\bibinfo {title} {{On a new
  method in the theory of superconductivity}},}\ }\href {\doibase
  10.1007/BF02745585} {\bibfield  {journal} {\bibinfo  {journal} {Il Nuovo
  Cimento}\ }\textbf {\bibinfo {volume} {7}},\ \bibinfo {pages} {794} (\bibinfo
  {year} {1958})}\BibitemShut {NoStop}%
\bibitem [{\citenamefont {Valatin}(1958)}]{Valatin1958}%
  \BibitemOpen
  \bibfield  {author} {\bibinfo {author} {\bibfnamefont {J.~G.}\ \bibnamefont
  {Valatin}},\ }\bibfield  {title} {\enquote {\bibinfo {title} {Comments on the
  theory of superconductivity},}\ }\href {\doibase 10.1007/BF02745589}
  {\bibfield  {journal} {\bibinfo  {journal} {Il Nuovo Cimento}\ }\textbf
  {\bibinfo {volume} {7}},\ \bibinfo {pages} {843} (\bibinfo {year}
  {1958})}\BibitemShut {NoStop}%
\bibitem [{\citenamefont {de~Gennes}\ and\ \citenamefont
  {Saint-James}(1963)}]{deGennes}%
  \BibitemOpen
  \bibfield  {author} {\bibinfo {author} {\bibfnamefont {P.~G.}\ \bibnamefont
  {de~Gennes}}\ and\ \bibinfo {author} {\bibfnamefont {D.}~\bibnamefont
  {Saint-James}},\ }\bibfield  {title} {\enquote {\bibinfo {title} {Elementary
  excitations in the vicinity of a normal metal-superconducting metal
  contact},}\ }\href {\doibase https://doi.org/10.1016/0031-9163(63)90148-3}
  {\bibfield  {journal} {\bibinfo  {journal} {Physics Letters}\ }\textbf
  {\bibinfo {volume} {4}},\ \bibinfo {pages} {151} (\bibinfo {year}
  {1963})}\BibitemShut {NoStop}%
\bibitem [{\citenamefont {Leggett}(2001)}]{LeggettRMP}%
  \BibitemOpen
  \bibfield  {author} {\bibinfo {author} {\bibfnamefont {Anthony~J.}\
  \bibnamefont {Leggett}},\ }\bibfield  {title} {\enquote {\bibinfo {title}
  {{Bose-Einstein condensation in the alkali gases: Some fundamental
  concepts}},}\ }\href {\doibase 10.1103/RevModPhys.73.307} {\bibfield
  {journal} {\bibinfo  {journal} {Rev. Mod. Phys.}\ }\textbf {\bibinfo {volume}
  {73}},\ \bibinfo {pages} {307--356} (\bibinfo {year} {2001})}\BibitemShut
  {NoStop}%
\bibitem [{\citenamefont {Kurita}\ \emph {et~al.}(2009)\citenamefont {Kurita},
  \citenamefont {Kobayashi}, \citenamefont {Morinari}, \citenamefont
  {Tsubota},\ and\ \citenamefont {Ishihara}}]{Kurita}%
  \BibitemOpen
  \bibfield  {author} {\bibinfo {author} {\bibfnamefont {Yasunari}\
  \bibnamefont {Kurita}}, \bibinfo {author} {\bibfnamefont {Michikazu}\
  \bibnamefont {Kobayashi}}, \bibinfo {author} {\bibfnamefont {Takao}\
  \bibnamefont {Morinari}}, \bibinfo {author} {\bibfnamefont {Makoto}\
  \bibnamefont {Tsubota}}, \ and\ \bibinfo {author} {\bibfnamefont {Hideki}\
  \bibnamefont {Ishihara}},\ }\bibfield  {title} {\enquote {\bibinfo {title}
  {{Spacetime analog of Bose-Einstein condensates: Bogoliubov--de Gennes
  formulation}},}\ }\href {\doibase 10.1103/PhysRevA.79.043616} {\bibfield
  {journal} {\bibinfo  {journal} {Phys. Rev. A}\ }\textbf {\bibinfo {volume}
  {79}},\ \bibinfo {pages} {043616} (\bibinfo {year} {2009})}\BibitemShut
  {NoStop}%
\bibitem [{\citenamefont {Hasan}\ and\ \citenamefont {Kane}(2010)}]{Hasan}%
  \BibitemOpen
  \bibfield  {author} {\bibinfo {author} {\bibfnamefont {M.~Z.}\ \bibnamefont
  {Hasan}}\ and\ \bibinfo {author} {\bibfnamefont {C.~L.}\ \bibnamefont
  {Kane}},\ }\bibfield  {title} {\enquote {\bibinfo {title} {{Colloquium:
  Topological insulators}},}\ }\href {\doibase 10.1103/RevModPhys.82.3045}
  {\bibfield  {journal} {\bibinfo  {journal} {Rev. Mod. Phys.}\ }\textbf
  {\bibinfo {volume} {82}},\ \bibinfo {pages} {3045--3067} (\bibinfo {year}
  {2010})}\BibitemShut {NoStop}%
\bibitem [{\citenamefont {Shams}\ \emph {et~al.}(2006)\citenamefont {Shams},
  \citenamefont {DuBois},\ and\ \citenamefont {Glyde}}]{Shams}%
  \BibitemOpen
  \bibfield  {author} {\bibinfo {author} {\bibfnamefont {Ali}\ \bibnamefont
  {Shams}}, \bibinfo {author} {\bibfnamefont {J.~L.}\ \bibnamefont {DuBois}}, \
  and\ \bibinfo {author} {\bibfnamefont {H.~R.}\ \bibnamefont {Glyde}},\
  }\bibfield  {title} {\enquote {\bibinfo {title} {{Localization of
  Bose--Einstein Condensation by Disorder}},}\ }\href {\doibase
  10.1007/s10909-006-9227-3} {\bibfield  {journal} {\bibinfo  {journal}
  {Journal of Low Temperature Physics}\ }\textbf {\bibinfo {volume} {145}},\
  \bibinfo {pages} {357} (\bibinfo {year} {2006})}\BibitemShut {NoStop}%
\bibitem [{\citenamefont {Griffin}\ and\ \citenamefont
  {Stringari}(1996)}]{Griffin}%
  \BibitemOpen
  \bibfield  {author} {\bibinfo {author} {\bibfnamefont {A.}~\bibnamefont
  {Griffin}}\ and\ \bibinfo {author} {\bibfnamefont {S.}~\bibnamefont
  {Stringari}},\ }\bibfield  {title} {\enquote {\bibinfo {title} {{Surface
  region of superfluid helium as an inhomogeneous Bose-condensed gas}},}\
  }\href {\doibase 10.1103/PhysRevLett.76.259} {\bibfield  {journal} {\bibinfo
  {journal} {Phys. Rev. Lett.}\ }\textbf {\bibinfo {volume} {76}},\ \bibinfo
  {pages} {259} (\bibinfo {year} {1996})}\BibitemShut {NoStop}%
\bibitem [{\citenamefont {Anglin}(2001)}]{Anglin}%
  \BibitemOpen
  \bibfield  {author} {\bibinfo {author} {\bibfnamefont {J.~R.}\ \bibnamefont
  {Anglin}},\ }\bibfield  {title} {\enquote {\bibinfo {title} {{Local vortex
  generation and the surface mode spectrum of large Bose-Einstein
  condensates}},}\ }\href {\doibase 10.1103/PhysRevLett.87.240401} {\bibfield
  {journal} {\bibinfo  {journal} {Phys. Rev. Lett.}\ }\textbf {\bibinfo
  {volume} {87}},\ \bibinfo {pages} {240401} (\bibinfo {year}
  {2001})}\BibitemShut {NoStop}%
\bibitem [{\citenamefont {Kuznetsov}\ and\ \citenamefont
  {Turitsyn}(1988)}]{Kuznetsov}%
  \BibitemOpen
  \bibfield  {author} {\bibinfo {author} {\bibfnamefont {E.~A.}\ \bibnamefont
  {Kuznetsov}}\ and\ \bibinfo {author} {\bibfnamefont {S.~K.}\ \bibnamefont
  {Turitsyn}},\ }\bibfield  {title} {\enquote {\bibinfo {title} {Instability
  and collapse of solitons in media with a defocusing nonlinearity},}\
  }\href@noop {} {\bibfield  {journal} {\bibinfo  {journal} {Sov. Phys. JETP}\
  }\textbf {\bibinfo {volume} {67}},\ \bibinfo {pages} {1583} (\bibinfo {year}
  {1988})}\BibitemShut {NoStop}%
\bibitem [{\citenamefont {Gaunt}\ \emph {et~al.}(2013)\citenamefont {Gaunt},
  \citenamefont {Schmidutz}, \citenamefont {Gotlibovych}, \citenamefont
  {Smith},\ and\ \citenamefont {Hadzibabic}}]{Gaunt}%
  \BibitemOpen
  \bibfield  {author} {\bibinfo {author} {\bibfnamefont {Alexander~L.}\
  \bibnamefont {Gaunt}}, \bibinfo {author} {\bibfnamefont {Tobias~F.}\
  \bibnamefont {Schmidutz}}, \bibinfo {author} {\bibfnamefont {Igor}\
  \bibnamefont {Gotlibovych}}, \bibinfo {author} {\bibfnamefont {Robert~P.}\
  \bibnamefont {Smith}}, \ and\ \bibinfo {author} {\bibfnamefont {Zoran}\
  \bibnamefont {Hadzibabic}},\ }\bibfield  {title} {\enquote {\bibinfo {title}
  {{Bose-Einstein Condensation of Atoms in a Uniform Potential}},}\ }\href
  {\doibase 10.1103/PhysRevLett.110.200406} {\bibfield  {journal} {\bibinfo
  {journal} {Phys. Rev. Lett.}\ }\textbf {\bibinfo {volume} {110}},\ \bibinfo
  {pages} {200406} (\bibinfo {year} {2013})}\BibitemShut {NoStop}%
\bibitem [{\citenamefont {Pikhitsa}(1992)}]{Pikhitsa}%
  \BibitemOpen
  \bibfield  {author} {\bibinfo {author} {\bibfnamefont {P.~V.}\ \bibnamefont
  {Pikhitsa}},\ }\bibfield  {title} {\enquote {\bibinfo {title} {Surface
  excitations of a nonideal bose gas},}\ }\href {\doibase
  https://doi.org/10.1016/0921-4526(92)90017-M} {\bibfield  {journal} {\bibinfo
   {journal} {Physica B: Condensed Matter}\ }\textbf {\bibinfo {volume}
  {179}},\ \bibinfo {pages} {201} (\bibinfo {year} {1992})}\BibitemShut
  {NoStop}%
\bibitem [{\citenamefont {Reut}\ and\ \citenamefont {Fisher}(1971)}]{Reut}%
  \BibitemOpen
  \bibfield  {author} {\bibinfo {author} {\bibfnamefont {L.~S.}\ \bibnamefont
  {Reut}}\ and\ \bibinfo {author} {\bibfnamefont {I.~Z.}\ \bibnamefont
  {Fisher}},\ }\bibfield  {title} {\enquote {\bibinfo {title} {{Surface
  excitations in liquid He$^4$}},}\ }\href@noop {} {\bibfield  {journal}
  {\bibinfo  {journal} {Sov. Phys. JETP}\ }\textbf {\bibinfo {volume} {33}},\
  \bibinfo {pages} {981} (\bibinfo {year} {1971})}\BibitemShut {NoStop}%
\bibitem [{\citenamefont {Chen}\ \emph {et~al.}(1998)\citenamefont {Chen},
  \citenamefont {Chen},\ and\ \citenamefont {Huang}}]{Chen}%
  \BibitemOpen
  \bibfield  {author} {\bibinfo {author} {\bibfnamefont {Xiang-Jun}\
  \bibnamefont {Chen}}, \bibinfo {author} {\bibfnamefont {Zhi-De}\ \bibnamefont
  {Chen}}, \ and\ \bibinfo {author} {\bibfnamefont {Nian-Ning}\ \bibnamefont
  {Huang}},\ }\bibfield  {title} {\enquote {\bibinfo {title} {{A direct
  perturbation theory for dark solitons based on a complete set of the squared
  Jost solutions}},}\ }\href {http://stacks.iop.org/0305-4470/31/i=33/a=005}
  {\bibfield  {journal} {\bibinfo  {journal} {Journal of Physics A:
  Mathematical and General}\ }\textbf {\bibinfo {volume} {31}},\ \bibinfo
  {pages} {6929} (\bibinfo {year} {1998})}\BibitemShut {NoStop}%
\bibitem [{\citenamefont {Muryshev}\ \emph {et~al.}(1999)\citenamefont
  {Muryshev}, \citenamefont {van Linden van~den Heuvell},\ and\ \citenamefont
  {Shlyapnikov}}]{Muryshev}%
  \BibitemOpen
  \bibfield  {author} {\bibinfo {author} {\bibfnamefont {A.~E.}\ \bibnamefont
  {Muryshev}}, \bibinfo {author} {\bibfnamefont {H.~B.}\ \bibnamefont {van
  Linden van~den Heuvell}}, \ and\ \bibinfo {author} {\bibfnamefont {G.~V.}\
  \bibnamefont {Shlyapnikov}},\ }\bibfield  {title} {\enquote {\bibinfo {title}
  {Stability of standing matter waves in a trap},}\ }\href {\doibase
  10.1103/PhysRevA.60.R2665} {\bibfield  {journal} {\bibinfo  {journal} {Phys.
  Rev. A}\ }\textbf {\bibinfo {volume} {60}},\ \bibinfo {pages} {R2665}
  (\bibinfo {year} {1999})}\BibitemShut {NoStop}%
\bibitem [{\citenamefont {Takahashi}\ \emph {et~al.}(2015)\citenamefont
  {Takahashi}, \citenamefont {Kobayashi},\ and\ \citenamefont {Nitta}}]{Nambu}%
  \BibitemOpen
  \bibfield  {author} {\bibinfo {author} {\bibfnamefont {Daisuke~A.}\
  \bibnamefont {Takahashi}}, \bibinfo {author} {\bibfnamefont {Michikazu}\
  \bibnamefont {Kobayashi}}, \ and\ \bibinfo {author} {\bibfnamefont {Muneto}\
  \bibnamefont {Nitta}},\ }\bibfield  {title} {\enquote {\bibinfo {title}
  {{Nambu-Goldstone modes propagating along topological defects: Kelvin and
  ripple modes from small to large systems}},}\ }\href {\doibase
  10.1103/PhysRevB.91.184501} {\bibfield  {journal} {\bibinfo  {journal} {Phys.
  Rev. B}\ }\textbf {\bibinfo {volume} {91}},\ \bibinfo {pages} {184501}
  (\bibinfo {year} {2015})}\BibitemShut {NoStop}%
\bibitem [{\citenamefont {Diallo}\ \emph {et~al.}(2014)\citenamefont {Diallo},
  \citenamefont {Azuah}, \citenamefont {Abernathy}, \citenamefont {Taniguchi},
  \citenamefont {Suzuki}, \citenamefont {Bossy}, \citenamefont {Mulders},\ and\
  \citenamefont {Glyde}}]{Diallo}%
  \BibitemOpen
  \bibfield  {author} {\bibinfo {author} {\bibfnamefont {S.~O.}\ \bibnamefont
  {Diallo}}, \bibinfo {author} {\bibfnamefont {R.~T.}\ \bibnamefont {Azuah}},
  \bibinfo {author} {\bibfnamefont {D.~L.}\ \bibnamefont {Abernathy}}, \bibinfo
  {author} {\bibfnamefont {Junko}\ \bibnamefont {Taniguchi}}, \bibinfo {author}
  {\bibfnamefont {Masaru}\ \bibnamefont {Suzuki}}, \bibinfo {author}
  {\bibfnamefont {Jacques}\ \bibnamefont {Bossy}}, \bibinfo {author}
  {\bibfnamefont {N.}~\bibnamefont {Mulders}}, \ and\ \bibinfo {author}
  {\bibfnamefont {H.~R.}\ \bibnamefont {Glyde}},\ }\bibfield  {title} {\enquote
  {\bibinfo {title} {{Evidence for a Common Physical Origin of the Landau and
  BEC Theories of Superfluidity}},}\ }\href {\doibase
  10.1103/PhysRevLett.113.215302} {\bibfield  {journal} {\bibinfo  {journal}
  {Phys. Rev. Lett.}\ }\textbf {\bibinfo {volume} {113}},\ \bibinfo {pages}
  {215302} (\bibinfo {year} {2014})}\BibitemShut {NoStop}%
\bibitem [{\citenamefont {Prisk}\ \emph {et~al.}(2013)\citenamefont {Prisk},
  \citenamefont {Das}, \citenamefont {Diallo}, \citenamefont {Ehlers},
  \citenamefont {Podlesnyak}, \citenamefont {Wada}, \citenamefont {Inagaki},\
  and\ \citenamefont {Sokol}}]{Prisk}%
  \BibitemOpen
  \bibfield  {author} {\bibinfo {author} {\bibfnamefont {Timothy~R.}\
  \bibnamefont {Prisk}}, \bibinfo {author} {\bibfnamefont {Narayan~C.}\
  \bibnamefont {Das}}, \bibinfo {author} {\bibfnamefont {Souleymane~O.}\
  \bibnamefont {Diallo}}, \bibinfo {author} {\bibfnamefont {Georg}\
  \bibnamefont {Ehlers}}, \bibinfo {author} {\bibfnamefont {Andrey~A.}\
  \bibnamefont {Podlesnyak}}, \bibinfo {author} {\bibfnamefont {Nobuo}\
  \bibnamefont {Wada}}, \bibinfo {author} {\bibfnamefont {Shinji}\ \bibnamefont
  {Inagaki}}, \ and\ \bibinfo {author} {\bibfnamefont {Paul~E.}\ \bibnamefont
  {Sokol}},\ }\bibfield  {title} {\enquote {\bibinfo {title} {Phases of
  superfluid helium in smooth cylindrical pores},}\ }\href {\doibase
  10.1103/PhysRevB.88.014521} {\bibfield  {journal} {\bibinfo  {journal} {Phys.
  Rev. B}\ }\textbf {\bibinfo {volume} {88}},\ \bibinfo {pages} {014521}
  (\bibinfo {year} {2013})}\BibitemShut {NoStop}%
\bibitem [{\citenamefont {Ginzburg}\ and\ \citenamefont
  {Pitaevskii}(1958)}]{Ginzburg1}%
  \BibitemOpen
  \bibfield  {author} {\bibinfo {author} {\bibfnamefont {V.~L.}\ \bibnamefont
  {Ginzburg}}\ and\ \bibinfo {author} {\bibfnamefont {L.~P.}\ \bibnamefont
  {Pitaevskii}},\ }\bibfield  {title} {\enquote {\bibinfo {title} {On the
  theory of superfluidity},}\ }\href {http://dx.doi.org/} {\bibfield  {journal}
  {\bibinfo  {journal} {Soviet Physics JETP}\ }\textbf {\bibinfo {volume}
  {7}},\ \bibinfo {pages} {858} (\bibinfo {year} {1958})}\BibitemShut {NoStop}%
\bibitem [{\citenamefont {Dziarmaga}(2004)}]{Dziarmaga}%
  \BibitemOpen
  \bibfield  {author} {\bibinfo {author} {\bibfnamefont {J.}~\bibnamefont
  {Dziarmaga}},\ }\bibfield  {title} {\enquote {\bibinfo {title} {{Quantum dark
  soliton: Nonperturbative diffusion of phase and position}},}\ }\href
  {\doibase 10.1103/PhysRevA.70.063616} {\bibfield  {journal} {\bibinfo
  {journal} {Phys. Rev. A}\ }\textbf {\bibinfo {volume} {70}},\ \bibinfo
  {pages} {063616} (\bibinfo {year} {2004})}\BibitemShut {NoStop}%
\bibitem [{Note1()}]{Note1}%
  \BibitemOpen
  \bibinfo {note} {The snake instability amounts to a {\protect \em moving}
  wall (a nodal plane) with its transverse parts moving at different
  velocities, which is hence acting to destroy the wall}\BibitemShut {NoStop}%
\bibitem [{\citenamefont {Landau}\ and\ \citenamefont
  {Lifshitz}(1977)}]{Landau}%
  \BibitemOpen
  \bibfield  {author} {\bibinfo {author} {\bibfnamefont {L.~D.}\ \bibnamefont
  {Landau}}\ and\ \bibinfo {author} {\bibfnamefont {E.~M.}\ \bibnamefont
  {Lifshitz}},\ }\href@noop {} {\emph {\bibinfo {title} {Quantum Mechanics}}}\
  (\bibinfo  {publisher} {Pergamon Press},\ \bibinfo {address} {New York},\
  \bibinfo {year} {1977})\ \bibinfo {note} {[p. 73, problem 5]}\BibitemShut
  {NoStop}%
\bibitem [{Note2()}]{Note2}%
  \BibitemOpen
  \bibinfo {note} {By making use of the known solutions of the homogeneous
  equations (\ref {eq:z1})~and~(\ref {eq:z2}) to satisfy the boundary
  conditions}\BibitemShut {NoStop}%
\bibitem [{\citenamefont {Ishkhanyan}(2015)}]{Ishkhanyan_2015}%
  \BibitemOpen
  \bibfield  {author} {\bibinfo {author} {\bibfnamefont {A.~M.}\ \bibnamefont
  {Ishkhanyan}},\ }\bibfield  {title} {\enquote {\bibinfo {title} {{Exact
  solution of the Schr{\"o}dinger equation for the inverse square root
  potential $V_0/\sqrt{x} $}},}\ }\href {\doibase 10.1209/0295-5075/112/10006}
  {\bibfield  {journal} {\bibinfo  {journal} {{EPL} (Europhysics Letters)}\
  }\textbf {\bibinfo {volume} {112}},\ \bibinfo {pages} {10006} (\bibinfo
  {year} {2015})}\BibitemShut {NoStop}%
\bibitem [{\citenamefont {Cotfas}\ and\ \citenamefont {Cotfas}(2011)}]{Cotfas}%
  \BibitemOpen
  \bibfield  {author} {\bibinfo {author} {\bibfnamefont {Nicolae}\ \bibnamefont
  {Cotfas}}\ and\ \bibinfo {author} {\bibfnamefont {Liviu~Adrian}\ \bibnamefont
  {Cotfas}},\ }\bibfield  {title} {\enquote {\bibinfo {title} {Hypergeometric
  type operators and their supersymmetric partners},}\ }\href {\doibase
  10.1063/1.3582829} {\bibfield  {journal} {\bibinfo  {journal} {Journal of
  Mathematical Physics}\ }\textbf {\bibinfo {volume} {52}},\ \bibinfo {pages}
  {052101} (\bibinfo {year} {2011})}\BibitemShut {NoStop}%
\bibitem [{\citenamefont {Tirao}(2003)}]{Tirao}%
  \BibitemOpen
  \bibfield  {author} {\bibinfo {author} {\bibfnamefont {Juan~A.}\ \bibnamefont
  {Tirao}},\ }\bibfield  {title} {\enquote {\bibinfo {title} {The matrix-valued
  hypergeometric equation},}\ }\href {\doibase 10.1073/pnas.1337650100}
  {\bibfield  {journal} {\bibinfo  {journal} {Proceedings of the National
  Academy of Sciences}\ }\textbf {\bibinfo {volume} {100}},\ \bibinfo {pages}
  {8138} (\bibinfo {year} {2003})}\BibitemShut {NoStop}%
\bibitem [{\citenamefont {J{\'o}dar}\ and\ \citenamefont
  {Cort{\'e}s}(1998)}]{Jodar}%
  \BibitemOpen
  \bibfield  {author} {\bibinfo {author} {\bibfnamefont {L}~\bibnamefont
  {J{\'o}dar}}\ and\ \bibinfo {author} {\bibfnamefont {J.C}\ \bibnamefont
  {Cort{\'e}s}},\ }\bibfield  {title} {\enquote {\bibinfo {title} {{Some
  properties of Gamma and Beta matrix functions}},}\ }\href {\doibase
  https://doi.org/10.1016/S0893-9659(97)00139-0} {\bibfield  {journal}
  {\bibinfo  {journal} {Applied Mathematics Letters}\ }\textbf {\bibinfo
  {volume} {11}},\ \bibinfo {pages} {89} (\bibinfo {year} {1998})}\BibitemShut
  {NoStop}%
\bibitem [{\citenamefont {Tamm}(1932)}]{Tamm}%
  \BibitemOpen
  \bibfield  {author} {\bibinfo {author} {\bibfnamefont {I.~E.}\ \bibnamefont
  {Tamm}},\ }\bibfield  {title} {\enquote {\bibinfo {title} {On the possible
  bound states of electrons on a crystal surface},}\ }\href@noop {} {\bibfield
  {journal} {\bibinfo  {journal} {Phys. Z. Sowjetunion}\ }\textbf {\bibinfo
  {volume} {1}},\ \bibinfo {pages} {733} (\bibinfo {year} {1932})}\BibitemShut
  {NoStop}%
\bibitem [{\citenamefont {Shockley}(1939)}]{Shockley}%
  \BibitemOpen
  \bibfield  {author} {\bibinfo {author} {\bibfnamefont {William}\ \bibnamefont
  {Shockley}},\ }\bibfield  {title} {\enquote {\bibinfo {title} {{On the
  Surface States Associated with a Periodic Potential}},}\ }\href {\doibase
  10.1103/PhysRev.56.317} {\bibfield  {journal} {\bibinfo  {journal} {Phys.
  Rev.}\ }\textbf {\bibinfo {volume} {56}},\ \bibinfo {pages} {317} (\bibinfo
  {year} {1939})}\BibitemShut {NoStop}%
\bibitem [{\citenamefont {Fischer}(2006)}]{Fischer}%
  \BibitemOpen
  \bibfield  {author} {\bibinfo {author} {\bibfnamefont {Uwe~R.}\ \bibnamefont
  {Fischer}},\ }\bibfield  {title} {\enquote {\bibinfo {title} {{Stability of
  quasi-two-dimensional Bose-Einstein condensates with dominant dipole-dipole
  interactions}},}\ }\href {\doibase 10.1103/PhysRevA.73.031602} {\bibfield
  {journal} {\bibinfo  {journal} {Phys. Rev. A}\ }\textbf {\bibinfo {volume}
  {73}},\ \bibinfo {pages} {031602} (\bibinfo {year} {2006})}\BibitemShut
  {NoStop}%
\bibitem [{\citenamefont {Kane}\ and\ \citenamefont {Lubensky}(2013)}]{Kane}%
  \BibitemOpen
  \bibfield  {author} {\bibinfo {author} {\bibfnamefont {C.~L.}\ \bibnamefont
  {Kane}}\ and\ \bibinfo {author} {\bibfnamefont {T.~C.}\ \bibnamefont
  {Lubensky}},\ }\bibfield  {title} {\enquote {\bibinfo {title} {Topological
  boundary modes in isostatic lattices},}\ }\href
  {https://doi.org/10.1038/nphys2835} {\bibfield  {journal} {\bibinfo
  {journal} {Nature Physics}\ }\textbf {\bibinfo {volume} {10}},\ \bibinfo
  {pages} {39} (\bibinfo {year} {2013})}\BibitemShut {NoStop}%
\end{thebibliography}%

\end{document}